# Toward Recovering Complete SRS for Softbody Simulation System and a Sample Application- a Team 9a SOEN6481 W13 Project Report

*Mahin Abbasipour (mah_abb@encs.concordia.ca)*

**Supervised By: Serguei A. Mokhov**

April 2013

# Contents

# List of Figures

# List of Tables

# Vision Document
# Softbody Simulation System

## 1. Introduction

This document aims at specifying the requirements and capturing the needs of users for building a softbody simulation system. This system has different applications ranging from computer games to surgery training which facilitates the creation and visualization of a certain softbody object. It also allows users to interact with created object at real time. A softbody or deformable object is an object whose shape changes due to an external force. Deformation type varies depending on the amount of object deformation. Each object can have multiple layers and each layer can have its own properties. So layers can be different in pressure, density and motion.

## 2. Positioning

### *2.1. Problem Statement*

| | |
|---|---|
| The problem of | • The existing softbody does not import a created object.<br>• There is no chance to apply new algorithms as plugin.<br>• It does not support fracture deformation. |
| Affects | Game makers, scientists, graphics researcher and end users who interact with softbody object. |
| The impact of which is | The combined objects with rigidbody cannot be created as the current system does not import rigid bodies. So it is not possible to attach a rigidbody to a softbody. |
| A successful solution would be | • The easy and automatic creation of a multi-layer softbody object.<br>• The easy changes of simulation parameters at run time.<br>• Allowing user to apply different algorithms and compare the efficiency of them. |

### *2.2. Product Position Statement*

| | |
|---|---|
| For | Game makers, scientists, graphics researcher and end users who interact with softbody object. |
| Who | Creates and simulates and does reseach |
| The Softbody Simulation System | Is a simulation framework. |
| That | Facilitates the easy creation and simulation of softbody object. |
| Unlike | Maya, Blender |
| Our product | Is open source. So it is good for educational purposes. User can interact with softbody and simulation is not predefined. Moreover, it is easy to work with. |

# 3. Stakeholder Descriptions

## *3.1. Stakeholder Summary*

This table shows the stakeholders of the softbody system and the responsibilities that they have.

| Name | Description | Responsibilities |
| --- | --- | --- |
| Graphics researcher, game makers | These stakeholders have basic knowledge for creation and modeling of a softbody object. | This stakeholder: creates the softbody object. |
| Market manager | This category of stakeholder knows the features the desired system should have and they know what the features of competitive products are. | This stakeholder: ensures that there will be a market demand for the product’s features |
| Developer | This category of stakeholder knows the needs of the users and features that the system-to-be should have. | This stakeholder: considers all requirements and develop a system that meets those requirements. |
| Project Manager | This category of stakeholder knows the system at high and abstract level. | This stakeholder: approves funding and monitor progress of the project. |
| Maintainer | This category of stakeholder knows the features of the system and user requirements. | This stakeholder: decides how a required and accepted change will affect the system. |
| Validator | This category of stakeholder knows both the needs of users and features of the system. | This stakeholder: Proves that the system meets the User Requirements (UR)s. |
| End users: kids, doctors, disabled people … | Depending on the type of application, the users may or may not have knowledge of how to use a computer system. | These stakeholders: interact with the system directly. |

## *3.2. User Environment*

For simulating the softbody system, depending on the simulation, various applications or devices may be needed to collaborate with the system. The platform for the current system can be Windows, OS X, Linux and Unix. Access to the system can be remotely or locally. For example, a group of students in a class remotely access a design model. Whereas a softbody model which helps conduct surgeries for doctors is used locally. The system may require necessary hardware.

# 4. Product Overview

## *4.1. Product Perspective*

This system consists of three separated components which are model, view and controller. Each of these components provides different functionalities. The model part creates the softbody according to the parameters that user gives to it. Controller component applies the forces to the object and calculates the next state of the object. The view component visualizes the simulation process. Other systems can use each component with collaboration of their system or users can use these three components together to create, simulate and visualize the simulation of a softbody object.

## *4.2. Assumptions and Dependencies*

| Assumptions | Dependencies |
|---|---|
| Various input/output devices like webcam, microphone, and touch screen… depending on the application are needed. | User to be able to interact with the created softbody object, input and output devices are needed. |

## *4.3. Needs and Features*

| Need | Need Description | Priority | Features | Feature Descriptions | Planned Release |
|---|---|---|---|---|---|
| **N01** Create softbody object | User should be able to create a multi-layer softbody object. | High | **F011** Provide default softbody object | The system should be able to create a default softbody object. | Iteration 2 |
| | | | **F012** Set initial parameters | The system should be able to take and set the density, dimension, size, mass, deformation range and texture of the softbody object. | Iteration 3 |
| | | | **F013** Add spring and face | The system should be able add particles by spring. The system should add face for 2D and 3D object. | Iteration 1 |
| | | | **F014** Attach softbody object to | The system should be able to attach the new created softbody object to the previously created objects if there is | Iteration 1 |
| | | | another object | any. | |
| **N02** Render softbody object | User should be able to view the created softbody object. | High | **F021** Create environment | The system should be able to draw the default or imported environment. | Iteration 1 |
| | | | **F022** Paint with color | The system should be able to draw the created softbody object inside a default or customized location and it is needed to render the object. | Iteration 1 |
| | | | **F023** Rotate softbody object | The system should be able to rotate softbody object after system draws it to provide different views. | Iteration 3 |
| | | | **F024** Set the angle of light | The system should be able to take and set the angle of light. | Iteration 2 |
| | | | **F025** provide default cameras | The system should be able to provide cameras with default position and orientation. | Iteration 2 |
| | | | **F026** set cameras | The system should be able to take and set the number of cameras, position and orientation of them. | Iteration 2 |
| | | | **F027** Enable or disable shader | The system should allow parameterized loading, compilation, linking and enabling/disabling of the shader. | Iteration 2 |
| **N03** Simulate softbody object | User should be able to interact with visualized softbody and observe its behavior at real time. | High | **F031** Apply algorithm | The system should be able to add multiple existing algorithms for Integration and collision detection. | Iteration 1 |
| | | | **F032** Change algorithm | The system should be able to change the algorithm at run time. | Iteration 3 |
| | | | **F033** process idle behavior | The system should be able to process the idle behavior of the softbody object. | Iteration 1 |
| **N04** Multiple displays | User should be able to add new display window. | Low | **F041** Add window | The system should be able to add multiple windows at simulation time either to display different views of the softbody object with the same applied algorithm or to display the same view of softbody object with different applied algorithms according to the angles of cameras. | Iteration 3 |
| **N05** Set simulation parameters | User should be able to set the parameters with new | High | **F051** Set Simulation parameters | The system should be able to take and set parameters like the no. of particles, mass of each particle, velocity and acceleration with new values at run time. | Iteration 1 |
| | values for softbody object. | | | | |
| **N06** Add new algorithm as plugin | User should be able to add new algorithm to the system as plugin | Low | **F061** Add new algorithm | The system should be able to take new algorithm as plugin either for integration or collision detection. | Iteration 3 |
| **N07** Import a created object or single or series of states for a simulated object | User should be able to import a created object or series of states for an object. | Normal | **F071** Import created object | The system should be able to import an object that another system created before. | Iteration 2 |
| | | | **F072** Import a single state | The system should be able to import a single state of softbody object that this system or another system simulated before and start simulation at that specific time. | Iteration 3 |
| | | | **F073** Import series of states | The system should be able to generate animation for a given series of states. | Iteration 3 |
| **N08** Export single or series of softbody object state. | User should be able to export the created softbody object. | Normal | **F081** Save a single state | The system should be able to save a single state at a given time. | Iteration 3 |
| | | | **F082** Save series of states | The system should be able to save a series of states at a given interval time. | Iteration 3 |
| **N09** Input/output devices | User should be able to interact with softbody object by input/output devices | High | **F091** Support sensors | The system should be able to response to actions received by input devices. | Iteration 1 |
| **N10** Export components of system | User should be able to use the function and classes as plugin and adapt it into another system. | Normal | **F101** Export Components | The system should provide the export of its component as plugin for another system. | Iteration 2 |
| **N11** Resume or pause simulation | The system should be able to resume a paused or pause a running simulation. | Low | **F111** Resume Simulation | The system should be able to resume a paused simulation. | Iteration 3 |
| | | | **F112** Pause Simulation | The system should be able to pause a running simulation. | Iteration 3 |

### *4.4. Alternatives and Competition*

The alternative system to this solution is using Maya but the usability of this system is low, and this system does not support real time simulation. Another solution is using AnyBody Modeling System which supports interaction at real time. But it is a software system for simulating the mechanics of the live human body. The interaction is defined in terms of external forces that user may impose or as a set recorded motion data. By AnyBody the user can obtain results on individual muscle forces, joint forces and moments, metabolism, elastic energy in tendons, antagonistic muscle actions [6]. Another solution is Lagoa [7] which is particle based and is for simulating the cloths and fluids. Although it has algorithms for collision detection, it does not support real time simulation.

# 5. Other Product Requirements

## 5.1. NFR01- Usability

The system should be easy to work for modeling, visualizing of the softbody and interacting with it. For example, varying the parameters at runtime needs a GUI for that so user can inter the parameters easily or allowing user to add multiple displays for a simulation.

## 5.2. NFR021- Response Time

The system must response at a reasonable time. For some applications, response at real time is important, whereas for others are not.

## *5.3. NFR022- Resource Utilization*

Identifies how much resource is needed at a unit of time so that the system can operate. This non-functional requirement depends on the efficiency of algorithms. For a given time step, the more computation is needed, the more resources are needed for computation.

## *5.4. NFR03- Accuracy*

Identifies how the created softbody object and its behavior at simulation time are near to reality. Accuracy depends on LOD and time step. Less time step and more LOD lead to a more accurate softbody object.

## *5.5. NFR04- Portability*

The system should work on other platforms and operating systems.

## *5.6. NFR05- Interoperability*

The components of the system should be able to be added on the other systems as plugins.

## *5.7. NFR06- Stability*

The simulation shouldn't crash at simulation time. Stability depends on the size of time step. The smaller time step leads to a more stable system.

### *5.8. NFR07- Reliability*

There must be a close approximation between the created softbody object and real object both in shape and behavior and response to the actions in real time. And the system shouldn't fail frequently and the repair time should be reasonable.

# Supplementary Specification (Document)

## 1. Introduction

### *1.1.Purpose*

This supplementary specification aims at specifying the definition of terms and abbreviations that are frequently used during the requirements document for softbody simulation system. It also specifies the functional and non-functional requirements for a softbody simulation system while it specifies each non-functional requirement is related to which functionality and what is the relationship between those nonfunctional requirements.

### *1.2. Scope*

This simulation system is facilitating the creation, visualization of a softbody while it simulates the objects behavior. The main characteristic of this kind of simulation is that the simulation is not predefined like the animation and it is possible to interact at real time with the created object.

### *1.3. Definitions, Acronyms and Abbreviations*

| | |
|---|---|
| D | Dimension |
| GUI | Graphical User Interface |
| LOD | Level of Details |
| MVC | Model View Controller |
| fps | Frame per Second |
| OS | Operating System |
| AHP | Analytic Hierarchy Process |
| BSD | Berkeley Software Distribution |
| OpenGL | Open Graphics Library |
| MTBF | Mean Time between Failures |
| MTTR | Mean Time to Repair |
| SRS | System Requirement Specification |

### *1.4. Overview*

The rest of supplementary document is organized as follow:

In section 2, functional requirements have been described. In section 3 to 11, the non-functional requirements have been specified. In section 11 and 12, design constraints and interfaces are described and the last section includes the glossary of terms.

# 2. Functionality

The list of features that captures the functional requirements is given in vision document in section 4.3, table of needs and features. This section describes the functional requirements of the softbody simulation system in natural language as follow:

- The system should be able to create a multi-layer softbody object.
- The system should be able to draw the softbody object.
- The system should be able to response to user's action at real time as feedback to the user.
- The system should be able to apply physics laws for all layers of the softbody object.
- The system should be able to take and apply different parameters as level of details for each layer to create the softbody object.
- The system should be able to accept changes in parameters at runtime without need to compile the simulation again.
- The system should be able to save a specific state of softbody object or a series of states at runtime, and export a saved state to different file formats like xml, txt, Excel Spreadsheet and sql.
- The system should be able to import a saved state and resume the simulation for it from that specific time.
- The system should be able to import a series of saved states and play animation for that.
- The system should make users able to reuse the components of the system as plugin to another system by exporting them.
- It should make users able to compare the performance of different algorithms for a softbody object at runtime without need to recompile the simulation either by changing the algorithm or adding a new algorithm at run time.
- The system should support multiple displays of the softbody when user wants to see different views of the same softbody object.
- The system should support multiple displays of the softbody object when user selects multiple algorithms to run simulation.
- The system should make users able to add new algorithms as plugins to the system.
- The softbody object should react to the collision and support elastic, plastic and fracture deformations.

- The system should be able to import the output of other softbody systems so the user would be able to Modify/complete the existing softbody object and run the simulation.
- The system should support the attachment of a softbody object to a rigidbody.
- The system should provide stereoscopic effects.
- The system should provide interaction through haptic devices and sensors.
- The system should allow parameterized loading, compilation, linking and enabling/disabling of the shader.

# 3. NFR01- Usability

To make the system understood, learnt and used by the user, different GUIs and windows should be used. For example, an interface will allow users to enter and control the number of particles, mass of each particle. Multiple displays have been used to allow users to observe the simulation of the same object from different views, or they observe the simulation of an object when apply different algorithms simultaneously. This feature allows users to visually compare the performance of different algorithms and see which one crash sooner or later. This non-functional requirement is related to the use cases that make the use of windows and interfaces for entering the simulation parameters. These use cases are UC01, UC02, UC04, UC05, UC06, UC07 and UC08.

# 4. NFR02- Performance

### *4.1. NFR021- Response time*

The system must response at a reasonable time. For some applications, response at real time is important and is considered as a functional requirement, whereas for others are not and response time is considered as the quality. Then, the type of this requirement depends on the type of application. The efficiency of the algorithm affects the response time, and it is in conflict with Accuracy. The more the object has particles the more accurate the object is, but for the point of computation, it needs more calculation as the number of particles increase. The related use cases are UC01 and UC02.

## *4.2.NFR022- Resource Utilization*

Identifies how much resource is needed at a unit of time so that the system can operate. This non-functional requirement depends on the efficiency of algorithms. For a given time step, the more computation is needed, the more resources are needed for computation. This non-functional requirement is related to the features in the use cases UC01 and UC02.

# 5. Supportability

N/A

# 6. NFR03- Accuracy

This requirement identifies if the created softbody object and its behavior at simulation time are near to reality. Accuracy depends on LOD and time step. Less time step and more LOD lead to a more accurate softbody object. This non-functional requirement is in conflict with response time. Because accuracy and response time have relation and are in conflict with each other, so their affected use cases are the same. They are UC01, UC02, UC04, UC05, UC06, UC07 and UC08.

# 7. NFR04- Portability

The system should work on other platforms and operating systems.

# 8. NFR05 Interoperability

The components of the system should be able to be added on the other systems as plugins. Moreover exporting the states of the simulation affects the interoperability. So the affected use case is UC09.

# 9. NFR06- Stability

The system shouldn't crash at simulation time. Stability depends on the size of time step. The smaller time step leads to a more stable system. So the related use case is UC01.

# 10. NFR07- Reliability

There must be a close approximation between the created softbody object and real object both in shape and behavior and response to the actions in real time. The system shouldn't fail frequently. MTBF should be around 24 hours and MTTR should be around one hour. The related use cases are the same as related use cases for accuracy.

# 11. Design Constraints

## *11.1. MVC Architectural Design*

This elastic object simulation system has been designed and implemented according to the architectural pattern Model-View-Controller.

## *11.2. Using OpenGL*

OpenGL is used as the core library.

### *11.3. Programming language*

C++ is used as programming language.

## 12. Purchased Components

N/A

## 13. Interfaces

### *13.1. User Interfaces*

We need to consider a GUI to make users able to enter input parameters for the creation of the softbody and applying different Algorithms to it. Moreover, user will interact with the created softbody as well.

### *13.2. Hardware Interfaces*

To use input devices like touch screen monitor, camera, mouse, keyboard, microphone and haptic devices, we need various hardware interfaces.

### *13.3. Software Interfaces*

The system consists of three separated components including model, view and controller. Although these components provide their own services, they need communication and the service of other components to provide their functionalities. So there must be interfaces between the components to communicate. For implementation these three components and their interfaces, OpenGL and C++ have been used.

### *13.4. Communications Interfaces*

Depending on the application, we need interfaces between the core softbody system and external collaborating systems. Consider the application which is developed for deaf people participating in a conference. The core system is the softbody simulation system and the collaborating systems are databases, and speech recognition system. In this application, the speech that is given by a speaker is recognized by the speech recognition system and with the help of databases it is understood that a specific world or feeling is equivalent to which position of the softbody object. So a set of words will be changed to some gestures in the softbody object at real time.

## 14. Licensing Requirements

BSD Open Source License.

## 15. Legal, Copyright and Other Notices

The copy right belongs to Ms. Miao Song and Concordia University.

# 16. Applicable Standards

N/A

# 17. Glossary

| term | Definition |
| --- | --- |
| Elastic Deformation | By applying force, the shape of object changes and by removing the force, the shape returns. |
| Plastic deformation | The object shape is deformed when tension is applied and its shape is partially returned to its original form when the force is removed. |
| Fracture deformation | The object is permanently deformed when it is irreversibly bent, torn, or broken apart after the material has reached the end of the elastic deformation ranges. |
| Elastic object | The object whose shape can change by applied force. |
| Particle | Each object is considered as a set of particles that have physical properties and are connected by spring to each other |
| Spring | Spring connects two particles to each other. |
| Face | The three connected springs is called face. |
| Haptic devices | Devices that provide force, vibration or motion to the user according to the input [8]. |
| Collision | An event that an object collides with the created softbody object. |
| Collision Detection | An algorithm that computes the applied force to the softbody object when the softbody object collides with another object. |
| Force | The strength that is applied to an object. |
| Integrator | An algorithm that computes the speed and position of each particle in a softbody object in each interval. Each algorithm has its own interval. |
| Rigidbody | An object made up incompliant materials and experience small amplitude of deformation during interaction with their environment [9]. |

| Acronym | Acronym Definition |
|---|---|
| D | **Dimension:** each object can be 1D, 2D or 3D. |
| GUI | **Graphical User Interface:** the interface that user interacts with the system via it. |
| LOD | **Level of Details:** the details of a softbody object which are the number of particles dimension… |
| MVC | **Model View Controller:** design architecture with three components of model, view and controller. |
| fps | **Frame per second**: the rate at which consecutive images are produced. |
| OS X | Operating System by Apple Inc. |
| Windows | Operating System by Microsoft. |
| MTBF | **Mean Time between Failures:** The time between two consequence failures. |
| MTTR | **Mean Time to Repair:** Time that takes to repair the failure and recover the service. |
| SRS | **Software Requirements Specification:** an IEEE standard that describes the content and quality of a good software specification [10]. |

# Use cases

# Revision History

| Date | Rev. | Description | Author(s) |
|---|---|---|---|
| 22/03/2013 | 1.0 | The brief use cases were created. | Mahin Abbasipour |
| 07/04/2013 | 1.1 | The use cases were elaborated to fully-dressed use cases. | Mahin Abbasipour |
| 10/04/2013 | 1.2 | The use case Simulate Softbody Object was refined. | Mahin Abbasipour |
| 11/04/2013 | 1.3 | The use cases Attach Objects and Simulate Softbody Object were refined. | Mahin Abbasipour |
| 15/04/2013 | 1.4 | The order of use cases changed by importance. | Mahin Abbasipour |

## 1. Actor-Goal List

| Actor | Goal |
|---|---|
| Primary Actor: Graphics Researcher | The graphics researcher wants to create, visualize the behavior of created softbody object under different algorithms and parameters. |
| Primary Actor: Softbody Simulation System | When there is no interaction between user and system, the Softbody object is idle. However, system needs interaction with itself to simulate the idle behavior and update the idle state at each time step. |
| Supporting actor: Sensors and actuators, other simulation systems | Devices allows user to interact with the created softbody object at simulation time and by the collaboration with other simulation systems, the import to or export from the simulation softbody system would be possible. |

## 2. Use Case Context Diagram

In this section, we describe the functional requirement and use cases.

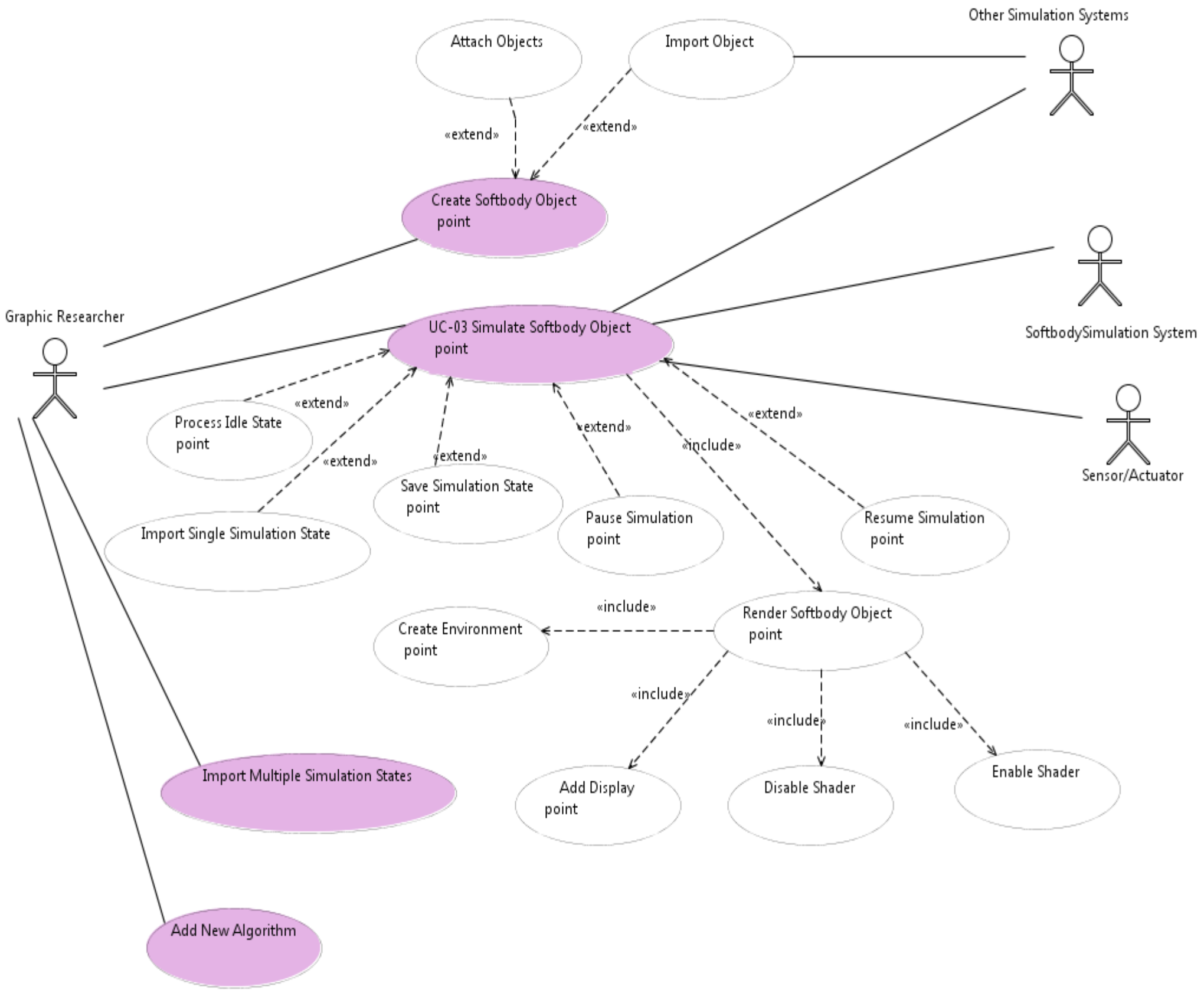

*Figure 1: use case context diagram for Softbody Simulation System*

## 3. Use case Model

This section includes three fully dressed use cases for simulating and creating the softbody object and attaching the created softbody object to other objects as follow:

**Id**: UC 01

**Use Case:** Simulate Softbody Object

**Description**
This use case describes the steps for visualizing the behavior of a created softbody.

**Level:** User Goal

**Primary Actor**

- Graphics researcher who interacts with softbody object.

**Supporting Actors**

- Sensors and actuators: by sensors system receives an action from user and system sends back its response to the user by actuators.
- Softbody Simulation System

**Stakeholders and Interests**

- Graphics researcher: wants to create and visualize the behavior of a softbody object.

**Pre-Conditions**

- User had created the softbody before.

**Post Conditions**

- System visualizes the behavior of softbody.

<u>Minimal Guarantee:</u>
System shows the idle behavior of softbody object.

## *Main Success Scenario*

1. User launches simulation execution.
2. User selects a created softbody object.
3. User adds deformation and integration algorithms.
   User repeats step 3 until user indicates done or user has added all of the existing algorithms.
4. System renders the softbody object.
5. User optionally interacts with the softbody object by sensor.
6. For each selected integrator algorithm, system calculates the drag force, collision force, pressure force, spring force and gravity force on each particle of softbody object.
7. System calculates the accumulated force on each particle.
8. System updates particles' property according to the accumulated force on each particle.
9. System updates the time step.
   System and user repeat steps 4 to 9.

## *Extensions*

2a. User imports a saved state to start simulation: Handle Import Single Simulation State.
2b. User imports a created softbody object from another softbody simulation system: Handle Import Object.

4a. Visualizing the created softbody object: Handle Render Softbody Object.
4b. System saves the state of object: Handle Save Simulation State.
4c. System visualizes the created softbody object and saves the state.
5-11a. User pauses the simulation: Handle Pause Simulation.
5-11b. User resumes the previously paused simulation: Handle Resume Simulation.

6-11a. User modifies the parameters of simulation:

1. System displays the parameters with previous values.
2. User modifies the values.
3. System draws the Softbody object with new parameters inside the environment.

6-11b. User replaces algorithms either for collision detection or integration:

1. System shows the algorithms that simulation is not running for it.
2. User selects an algorithm.
3. System applies the algorithm.
4. System updates the time step for algorithm.

8a. there is no drag or collision force: Handle Process Idle State.

## *Special Requirements*

- The efficiency of the algorithm user chooses for simulation affects the performance metrics.
- The parameters like the number of particles that user may modify during the simulation affects the accuracy of softbody object, and the number of computations that algorithm uses in each time step.
- The GUI for getting the simulation parameters affects the usability of the system.

---

**Id**: UC02

**Use Case:** Create Softbody Object

**Description**
This use case describes the steps for the creation of a softbody object.

**Level:** User Goal

**Primary Actor**
- Graphics researcher

**Supporting Actors**
- There is no supporting actor.

**Stakeholders and Interests**

- Graphics researcher: wants to create a softbody object.

**Pre-Conditions**
There is no specific pre-condition.

**Post Conditions**
The system will create and save a softbody object successfully.

Minimal Guarantee
System creates the default softbody object.

## *Main Success Scenario*

1. User opens a blank project.
2. System shows the default initial parameters like the number of particles, mass of each particle, dimension, density, color and texture of softbody object to create it.
3. User optionally modifies theses initial values.
4. System sets the particles.
5. System adds spring between two particles of a softbody object.
6. System adds face for 2D and 3D objects.
   System and user repeats steps 2 to 6 until user indicates done.
7. User saves the created softbody object.

## *Extensions*

1a. User opens a previously created project:
1. User chooses the path the project exists.
2. System fetches the project and loads it.

1b. User imports a softbody object that another simulation system created it before: Handle Import Object.
7a. User attaches the created objects to each other: Handle Attach Objects

## *Special Requirements*

- The Accuracy of created softbody object depends on the parameters like the number of particles that user entered.
- The GUI for giving the number of particles, mass of each particle … affects the usability of the system.

---

**Id**: UC03

**Use Case:** Attach Objects

**Description**
This use case describes the steps for attaching the created softbody object with other objects.

**Level:** Subfunction

**Primary Actor**

– Graphics researcher

**Supporting Actors**

– There is no specific supporting actor.

**Stakeholders and Interests**

– Graphics researcher: wants to create a softbody.

**Pre-Conditions**

– There must be at least two objects to attach to each other.

**Post Conditions**

– System connects objects and saves the combined object successfully.

<u>Minimal Guarantee</u>
There is no minimal guarantee.

### *Main Success Scenario*

1. User selects a set of particles from two different objects.
2. System adds spring between two selected set of particles.
   User and system repeats steps 1 and 2 until user indicates done or there is no pair to attach.
3. User saves the created object.

### *Extension*

1a. User selects particle from one object:

1. System shows an error message that selected particles should be from two different objects.

### *Special Requirements*

– The Accuracy of created softbody object depends on the parameters like the number of particles that user entered.

## 4. Brief Use Cases

This section includes a brief explanation of use cases mentioned in the context diagram as brief use cases.

**Id**:  UC04
**Use Case:** Render Softbody Object
**Actor:** Graphics researcher

In this use case, user at first sets the position and orientation of sources of light and system applies them. After that, user sets the orientation, location and number of cameras and specifies the camera for showing in the main window. System draws the softbody object with color inside the created environment in the main window. System renders softbody object after each time step. By adding display, system renders the softbody object. For rendering, user may or may not enable or disable the shader.

---

**Id**:  UC05
**Use Case:** Create Environment
**Actor:** Graphics researcher

For creating the environment, system imports the previously created environments and like the creation of softbody object, it does not take parameters from user to create it.

---

**Id:** UC06
**Use Case:** Add Display
**Actor:** Graphics researcher

User may choose to add window either for different views of a Softbody or for different algorithms that are running simultaneously. If user adds display for a different view, system renders and visualizes the behavior of Softbody object from the time system creates the new window for that specific view. If user chooses to add new display for added algorithm, after creating environment inside the new created main window, system creates a new instance of a simulation and softbody object and renders the new softbody object.

---

**Id:** UC07
**Use Case:** Pause Simulation
**Actor**: Graphics researcher

User pauses the simulation. For this purpose, temporarily system stops the simulation and saves the simulation parameters of the softbody object. System draws the last state the softbody object was.

---

**Id:** UC08

**Use Case:** Resume Simulation
**Actor:** Graphics researcher

For resuming a simulation, the system must be in paused state before. For this purpose, system fetches the parameters that system had saved before and continues simulation according to those parameters.

---

**Id:** UC09
**Use Case:** Save Simulation State
**Actor**: Graphics researcher

When the simulation is running, user can save a single state or a series of states. If it is single state, system saves the state of current time. Otherwise, if the simulation is a series of states, then system saves the states at each time step for an interval.

---

**Id:** UC10
**Use Case:** Import Object
**Primary Actor:** Graphics researcher
**Supporting Actor:** Other Simulation Systems

User may want to import an object (either softbody or rigidbody object) which another system created. For this purpose, user selects the path that object exists. System loads that project and fetches all the parameters for that object.

---

**Id:** UC11
**Use Case:** Enable Shader
**Actor:** Graphics researcher

User selects enabling the shader and system applies the changes in the softbody object.

---

**Id:** UC12
**Use Case:** Disable Shader
**Actor:** Graphics researcher

User selects disabling the shader and system applies the changes in the softbody object.

---

**Id:** UC13
**Use Case:** Process Idle State
**Actor:** Softbody Simulation System

When user is not interacting with softbody object or there is no external force like collision, for each particle of the softbody object, system updates the position and velocity of the particles according to the behavior of softbody when it is idle.

---

**Id:** UC14
**Use Case:** Import Single Simulation State
**Actor:** Graphics researcher

User may want to import a previously saved state to start simulation from that specific state. For this purpose, user selects the path the saved state exists. System loads that project and fetches all the parameters for simulation and simulation at real time starts. User may or may not interact with softbody object.

---

**Id:** UC15
**Use Case:** Import Multiple Simulation States
**Actor:** Graphics researcher

User may want to import a series of states to play a previously simulated softbody object. For this purpose, user selects the path the saved states exist. System fetches the parameters. User cannot interact with softbody object.

---

**Id:** UC16
**Use Case:** add New Algorithm
**Actor:** Graphics researcher

User may want to add a new algorithm as a plugin. User selects the algorithm. System adds the algorithm and algorithm exists as available integration or collision detection algorithm.

---

# 5. Cost value prioritization

For setting priorities among a set of requirements captured by use cases Create Softbody Object and Render Softbody Object, Analytic Hierarchy Process (AHP) method is used. Table 1 shows the comparison of requirements with each other according to the values they contribute to the objectives. These values are 1, 3, 5, 7 and 9 which are relatives [5]. These values were obtained from stakeholders. Table 2 is the normalized relative contribution of requirements. Similarly, tables 3 and 4 show the relative costs of requirements. The values for cost were obtained by estimation. For a full list of features refer to vision document, section 4.3, table for features and needs. The reason for not bringing the non-functional requirements in these comparisons is that non-functional requirements are related to a specific feature (i.e. functional requirement) and they don't exist by themselves alone.

| | F011 | F012 | F013 | F014 | F021 | F022 | F023 | F024 | F025 | F026 |
|---|---|---|---|---|---|---|---|---|---|---|
| F011 | **1** | 0.11111111 | 0.11111111 | 0.2 | 3 | 0.2 | 3 | 3 | 3 | 0.333333 |
| F012 | 9 | **1** | **3** | 5 | 9 | 7 | 9 | 9 | 9 | 7 |
| F013 | 9 | 0.33333333 | **1** | 7 | 9 | 9 | 9 | 9 | 9 | 7 |
| F014 | 5 | 0.2 | 0.14285714 | **1** | 7 | 5 | 7 | 7 | 7 | 5 |
| F021 | 0.33333333 | 0.11111111 | 0.11111111 | 0.142857 | **1** | 0.2 | 3 | 3 | 0.333333 | 0.2 |
| F022 | 5 | 0.14285714 | 0.11111111 | 0.2 | 5 | **1** | 7 | 5 | 7 | 3 |
| F023 | 0.33333333 | 0.11111111 | 0.11111111 | 0.142857 | 0.333333 | 0.142857 | **1** | 0.2 | 3 | 0.2 |
| F024 | 0.33333333 | 0.11111111 | 0.11111111 | 0.142857 | 0.333333 | 0.2 | 5 | **1** | 3 | 0.2 |
| F025 | 0.33333333 | 0.11111111 | 0.11111111 | 0.142857 | 3 | 0.142857 | 0.333333 | 0.33333333 | **1** | 0.2 |
| F026 | 3 | 0.14285714 | 0.14285714 | 0.2 | 5 | 0.333333 | 5 | 5 | 5 | **1** |
| Sum | 33.3333333 | 2.37460317 | 4.95238095 | 14.17143 | 42.66667 | 23.21905 | 49.33333 | 42.5333333 | 47.33333 | 24.13333 |

*Table 1: AHP comparison table with relative values of requirements*

| | F011 | F012 | F013 | F014 | F021 | F023 | F024 | F025 | F026 | F027 | RV |
|---|---|---|---|---|---|---|---|---|---|---|---|
| F011 | 0.03 | 0.04679144 | 0.0224359 | 0.014113 | 0.070313 | 0.008614 | 0.060811 | 0.07053292 | 0.06338 | 0.013812 | 0.04008 |
| F012 | 0.27 | 0.42112299 | 0.60576923 | 0.352823 | 0.210938 | 0.301477 | 0.182432 | 0.21159875 | 0.190141 | 0.290055 | 0.303636 |
| F013 | 0.27 | 0.14037433 | 0.20192308 | 0.493952 | 0.210938 | 0.387613 | 0.182432 | 0.21159875 | 0.190141 | 0.290055 | 0.257903 |
| F014 | 0.15 | 0.0842246 | 0.02884615 | 0.070565 | 0.164063 | 0.21534 | 0.141892 | 0.1645768 | 0.147887 | 0.207182 | 0.137458 |
| F021 | 0.01 | 0.04679144 | 0.0224359 | 0.010081 | 0.023438 | 0.008614 | 0.060811 | 0.07053292 | 0.007042 | 0.008287 | 0.026803 |
| F022 | 0.15 | 0.06016043 | 0.0224359 | 0.014113 | 0.117188 | 0.043068 | 0.141892 | 0.11755486 | 0.147887 | 0.124309 | 0.093861 |
| F023 | 0.01 | 0.04679144 | 0.0224359 | 0.010081 | 0.007812 | 0.006153 | 0.02027 | 0.00470219 | 0.06338 | 0.008287 | 0.019991 |
| F024 | 0.01 | 0.04679144 | 0.0224359 | 0.010081 | 0.007812 | 0.008614 | 0.101351 | 0.02351097 | 0.06338 | 0.008287 | 0.030226 |
| F025 | 0.01 | 0.04679144 | 0.0224359 | 0.010081 | 0.070313 | 0.006153 | 0.006757 | 0.00783699 | 0.021127 | 0.008287 | 0.020978 |
| F026 | 0.09 | 0.06016043 | 0.02884615 | 0.014113 | 0.117188 | 0.014356 | 0.101351 | 0.11755486 | 0.105634 | 0.041436 | 0.069064 |
| Sum | 1 | 1 | 1 | 1 | 1 | 1 | 1 | 1 | 1 | 1 | 1 |

*Table 2: AHP normalized table with relative contribution of requirements to the overall value of project*

| | F011 | F012 | F013 | F014 | F021 | F023 | F024 | F025 | F026 | F027 |
|---|---|---|---|---|---|---|---|---|---|---|
| F011 | **1** | 5 | 3 | 3 | 3 | 0.2 | 0.14285714 | 7 | 7 | 7 |
| F012 | 0.2 | **1** | 0.33333333 | 0.14285714 | 7 | 0.33333333 | 0.11111111 | 3 | 3 | 3 |
| F013 | 0.33333333 | 3 | **1** | 0.2 | 0.33333333 | 3 | 0.14285714 | 5 | 5 | 5 |
| F014 | 0.33333333 | 7 | 5 | **1** | 3 | 7 | 3 | 7 | 7 | 7 |
| F021 | 0.33333333 | 0.14285714 | 3 | 0.33333333 | **1** | 3 | 3 | 5 | 5 | 5 |
| F023 | 5 | 3 | 0.33333333 | 0.14285714 | 0.33333333 | **1** | 3 | 3 | 3 | 3 |
| F024 | 7 | 9 | 7 | 0.33333333 | 0.33333333 | 0.33333333 | **1** | 9 | 9 | 9 |
| F025 | 0.14285714 | 0.33333333 | 0.2 | 0.14285714 | 0.2 | 0.33333333 | 0.11111111 | **1** | 1 | 1 |
| F026 | 0.14285714 | 0.33333333 | 0.2 | 0.14285714 | 0.2 | 0.33333333 | 0.11111111 | 1 | **1** | 3 |
| F027 | 0.14285714 | 0.33333333 | 0.2 | 0.14285714 | 0.2 | 0.33333333 | 0.11111111 | 1 | 0.333333 | **1** |
| Sum | 14.6285714 | 29.1428571 | 20.2666667 | 5.58095238 | 15.6 | 15.8666667 | 10.7301587 | 42 | 41.33333 | 44 |

*Table 3: AHP comparison table with relative cost of requirements*

| | F011 | F012 | F013 | F014 | F021 | F023 | F024 | F025 | F026 | F027 | RV |
|---|---|---|---|---|---|---|---|---|---|---|---|
| F011 | 0.06835938 | 0.17156863 | 0.14802632 | 0.53754266 | 0.19230769 | 0.01260504 | 0.01331361 | 0.16666667 | 0.169355 | 0.15909091 | 0.16388357 |
| F012 | 0.01367188 | 0.03431373 | 0.01644737 | 0.02559727 | 0.44871795 | 0.0210084 | 0.01035503 | 0.07142857 | 0.072581 | 0.06818182 | 0.07823027 |
| F013 | 0.02278646 | 0.10294118 | 0.04934211 | 0.03583618 | 0.02136752 | 0.18907563 | 0.01331361 | 0.11904762 | 0.120968 | 0.11363636 | 0.07883144 |
| F014 | 0.02278646 | 0.24019608 | 0.24671053 | 0.17918089 | 0.19230769 | 0.44117647 | 0.2795858 | 0.16666667 | 0.169355 | 0.15909091 | 0.20970563 |
| F021 | 0.02278646 | 0.00490196 | 0.14802632 | 0.05972696 | 0.06410256 | 0.18907563 | 0.2795858 | 0.11904762 | 0.120968 | 0.11363636 | 0.11218574 |
| F023 | 0.34179688 | 0.10294118 | 0.01644737 | 0.02559727 | 0.02136752 | 0.06302521 | 0.2795858 | 0.07142857 | 0.072581 | 0.06818182 | 0.10629523 |
| F024 | 0.47851563 | 0.30882353 | 0.34539474 | 0.05972696 | 0.02136752 | 0.0210084 | 0.09319527 | 0.21428571 | 0.217742 | 0.20454545 | 0.19646051 |
| F025 | 0.00976563 | 0.01143791 | 0.00986842 | 0.02559727 | 0.01282051 | 0.0210084 | 0.01035503 | 0.02380952 | 0.024194 | 0.02272727 | 0.01715835 |
| F026 | 0.00976563 | 0.01143791 | 0.00986842 | 0.02559727 | 0.01282051 | 0.0210084 | 0.01035503 | 0.02380952 | 0.024194 | 0.06818182 | 0.02170381 |
| F027 | 0.00976563 | 0.01143791 | 0.00986842 | 0.02559727 | 0.01282051 | 0.0210084 | 0.01035503 | 0.02380952 | 0.008065 | 0.02272727 | 0.01554545 |
| sum | 1 | 1 | 1 | 1 | 1 | 1 | 1 | 1 | 1 | 1 | 1 |

*Table 4: AHP normalized table with relative contribution of requirements to the overall cost of project*

Figure 2 shows the value-cost prioritization for creating and rendering the softbody object. The x axis is the cost percentage and the y axis is the value percentage. As shown in this figure, on the whole, the priority for creating a softbody object is more than the priority for rendering the softbody object. However, some specific features like creating a default softbody object (F011) has less priority than drawing the softbody object inside the environment (F022).

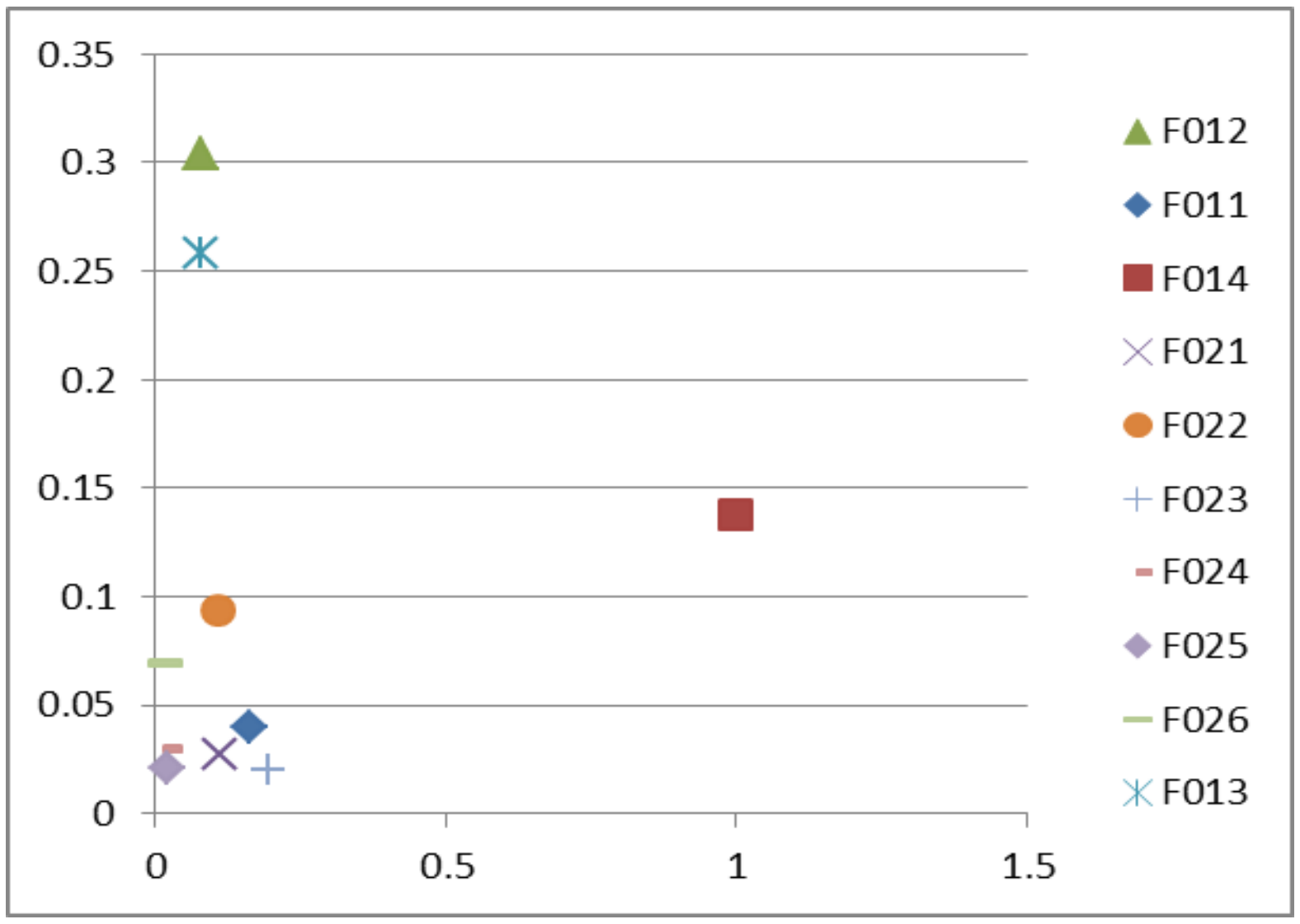

*Figure 2: value-cost requirements prioritization*

# Diagrams

## 1. Domain model

Figure 3 shows the concepts in softbody simulation system and relationship among them. In this section, the concepts, their attributes, potential constraints and the associations with other concepts are explained.

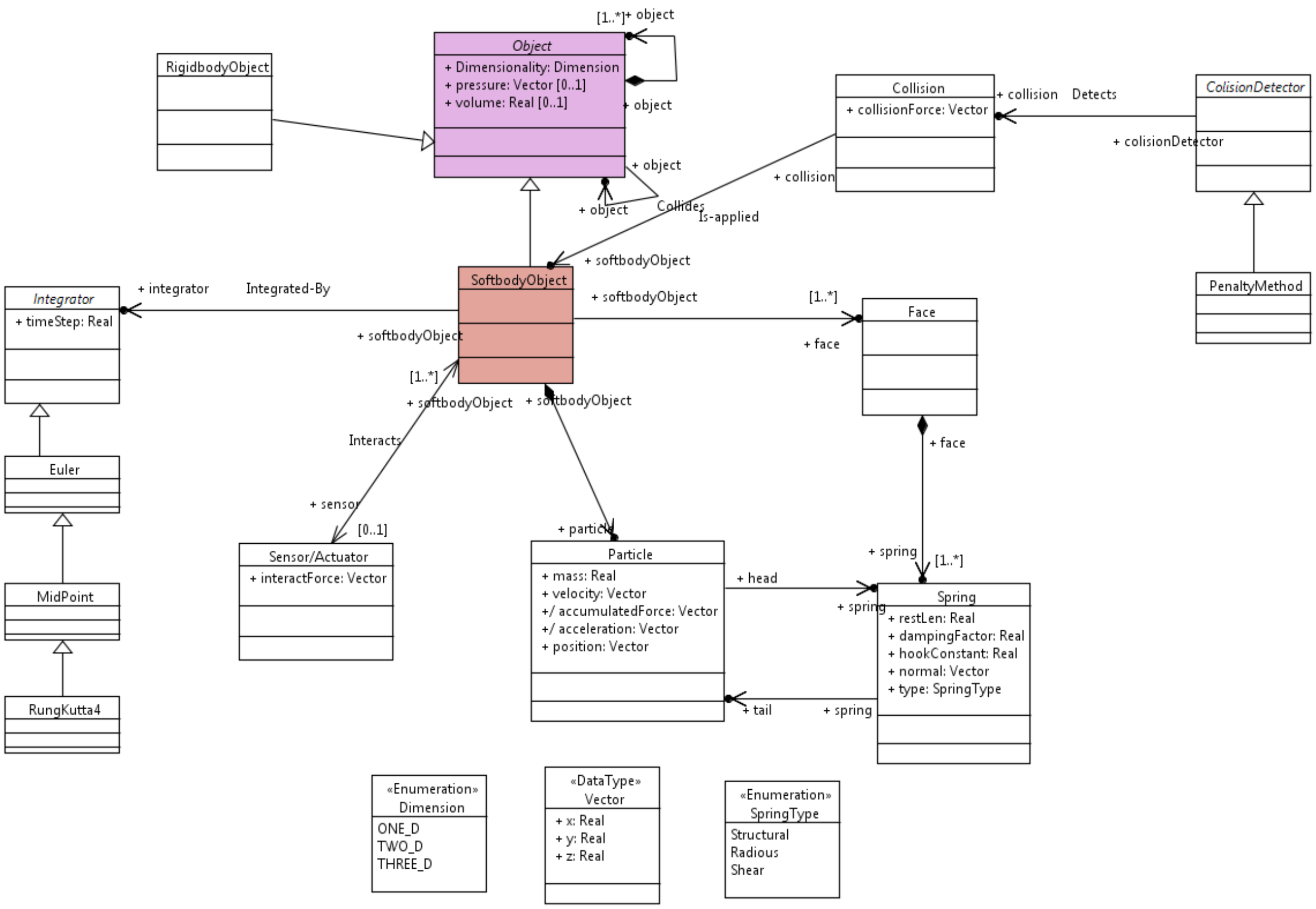

*Figure 3: domain model for softbody simulation system*

## *1.1 Object*

An object can be a softbody object or a regidbody object or the combination of both. This class is an abstract class.

### 1.1.1 Attributes

- dimensionality: this attribute shows the dimension of an object. This attribute can take one of the three values of ONE_D, TWO_D or THREE_D.

- pressure [0..1]: this attribute shows the pressure inside a 2-D or 3-D object. The multiplicity of this attribute which is shown by [0..1], must be zero for a 1-D object.
- volume [0..1]: this attribute shows the volume of a 3-D object.

### 1.1.2 Associations

- An object can collide with another object (a softbody object can collide with a rigidbody or another softbody object).
- An object may consist of multiple layers that are attached to each other which can be softbody or rigidbody.

### 1.1.3 Constraints

The multiplicity of attribute volume and pressure must be zero for a 1-D object and for a 2-D and 1-D object the multiplicity of attribute volume must be zero.

## *1.2 SoftbodyObject*

This class is the specialization of the abstract class Object.

### 1.2.1 Attributes

There is no specific attributes for this class.

### 1.2.2 Associations

A softbody object consists of one to many particles and zero to many faces. A softbody object receives user's action by sensors and sends the response by actuators. Moreover, each instance of SoftbodyObject integrated by only one integrator algorithm. A collision can be applied to a softbody which has a collision force.

### 1.2.3 Constraints

There must be a constraint that there is no face for a one-D object so in this case the multiplicity of association must be always zero.

## *1.3 Particle*

The smallest part of a softbody object is particle.

### 1.3.1 Attributes

- mass: this attribute shows the mass of each particle. So by the whole mass of particles consist a softbody object, the weight of softbody object can be obtained.
- velocity: each particle has a velocity. This attribute is the type vector because the velocity of each particle can be in different directions (x,y,z).
- acceleration: this attribute can be derived from velocity. The type of this attribute is vector.

- position: each particle of a softbody object has its own location inside the environment.
- accumulatedForce: this attribute shows the accumulated force applied on each particle.

### 1.3.2 Constraints

If the object is one-D the value for y and z of vector for velocity, acceleration, position and accumulatedForce must be set to zero. Similarly, for a 2-D object, the y value must be set to zero.

## *1.4 Spring*

Spring connects two particles to each other.

### 1.4.1 Attributes

- resLen: this attribute shows the length of spring when it is in rest.
- dampingFactor: this attribute shows the spring's damping factor.
- hookConstant: this attribute shows the spring's hook constant.
- normal: shows the spring's normal vector.
- type: this attribute shows the type of spring which can be structural, radius and shear.

### 1.4.2 Associations

This class has association with particle that shows the head and tail of spring.

### 1.4.3 Constraint

There is no specific constraint for this class.

## *1.5 Face*

Face consists of springs. This class has no specific attribute or constraint.

## *1.6 Integrator*

This abstract class is for representing different algorithms exists for integration. Each softbody is integrated by an algorithm. When an algorithm is needed to be added as a plugin to the system, it will come as the specialization of the class integrator.

### 1.6.1 Attributes

timeStep: this attribute shows the interval time that each algorithm does calculation for accumulated force on each particle.

### *1.7 CollisionDetector*

This abstract class is for classifying different types of algorithms exist for collision detection.

## 2. Sequence Diagram

Figures 4 and 5 show the sequence diagram for the most important functionalities of the system, Simulate Softbody Object (UC01) and Create Softbody Object (UC02). These sequence diagrams consider main scenarios only.

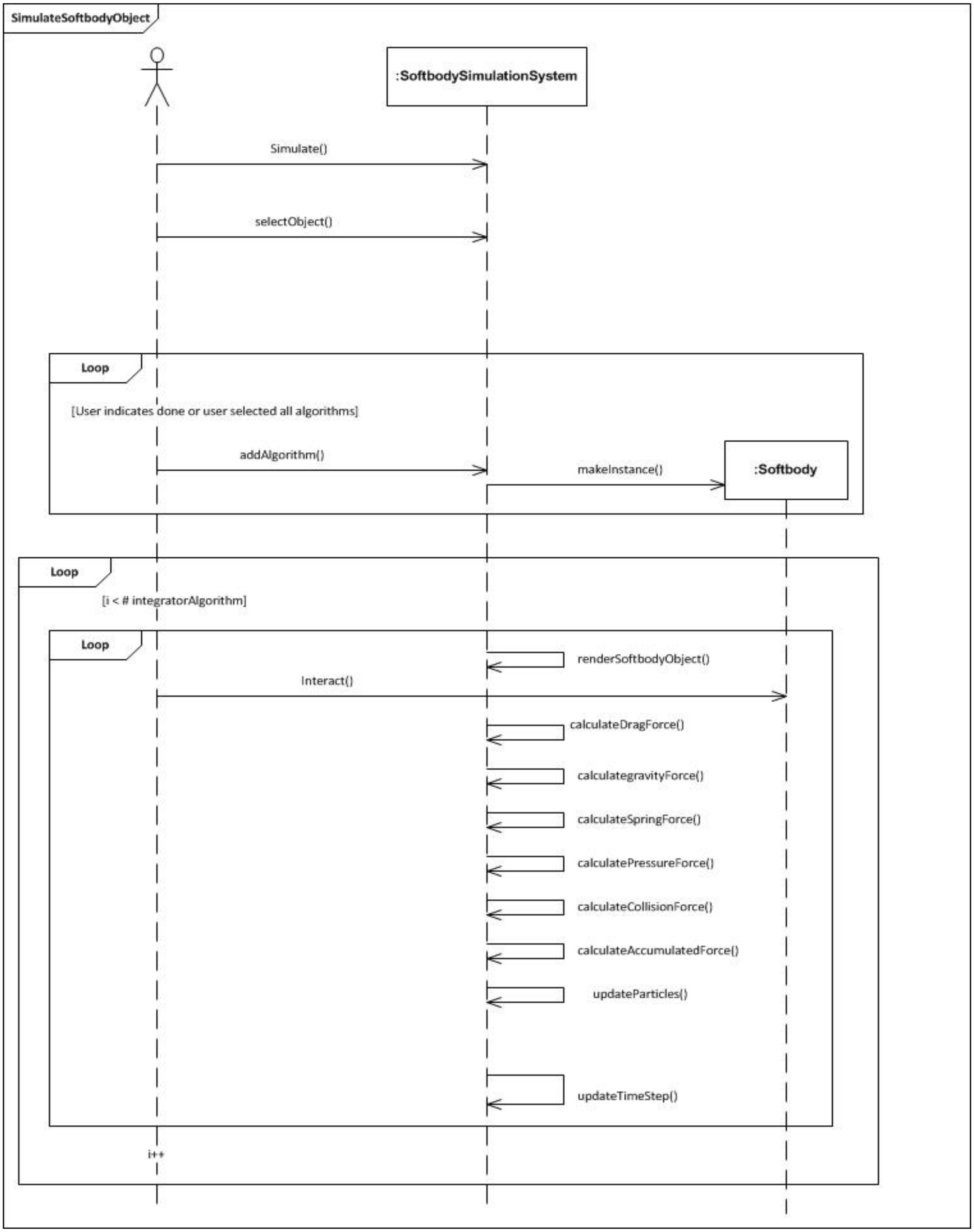

*Figure 4: sequence diagram for Simulate Softbody Object- main scenario*

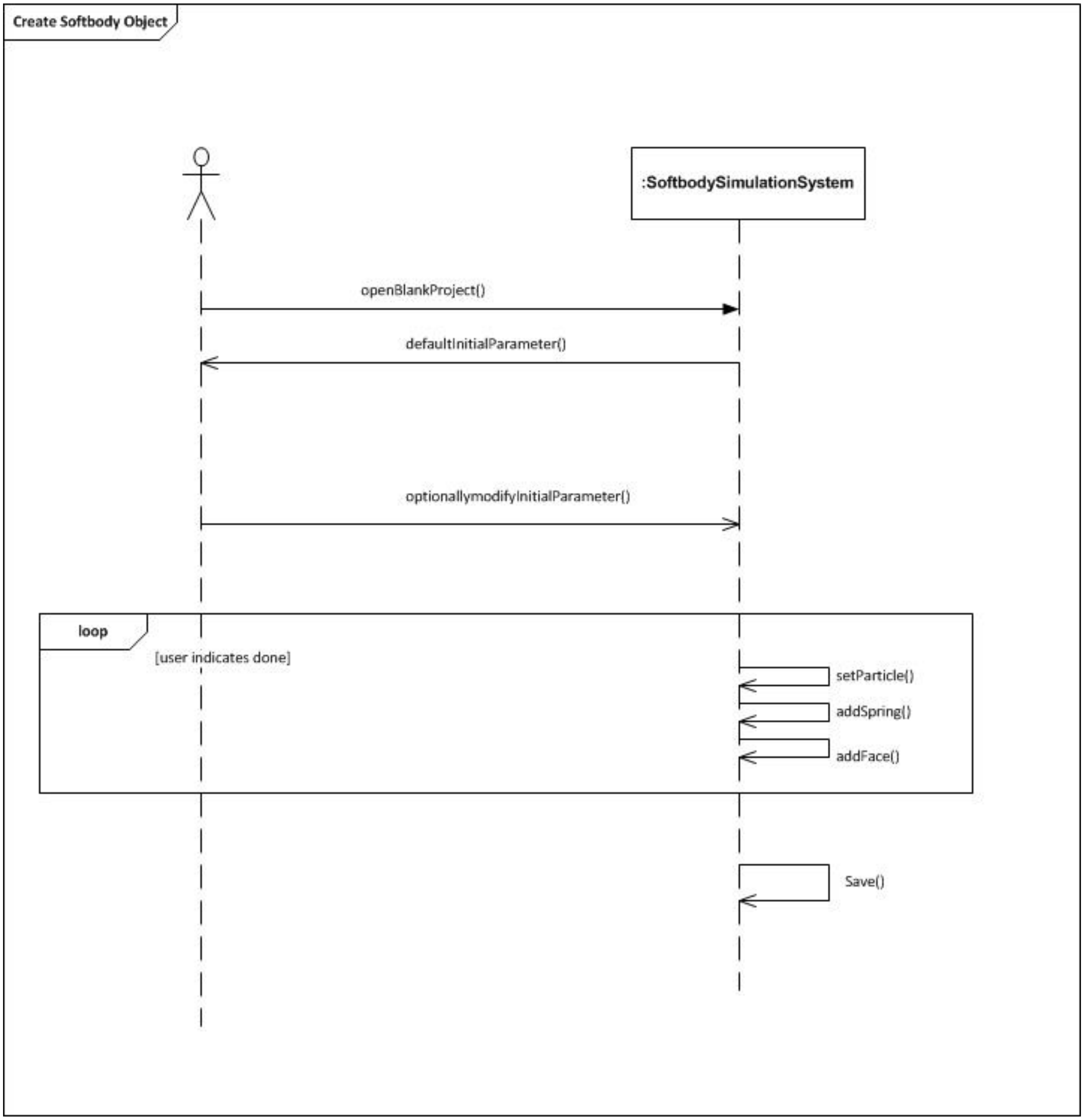

*Figure 5: sequence diagram for Create Softbody Object-main scenario*

# 3. Activity Diagram

This section shows the activity diagram for use cases Simulate Softbody Object (UC01) and Create Softbody Object (UC2) while it considers the alternative scenarios as well.
Figure 6 shows the activity diagram called SingleAlgo. This activity diagram shows the states which are a part of UC01. This diagram shows the steps when only one integrator algorithm is chosen. By each algorithm, a new instance of softbody object is created. At each time step, the accumulation of different forces is calculated by system and depending on what user chooses, the updated state is visually shown or saved or both at the same time. During these steps, user optionally can change one algorithm of a same simulation, change the parameters or interact with the softbody object.

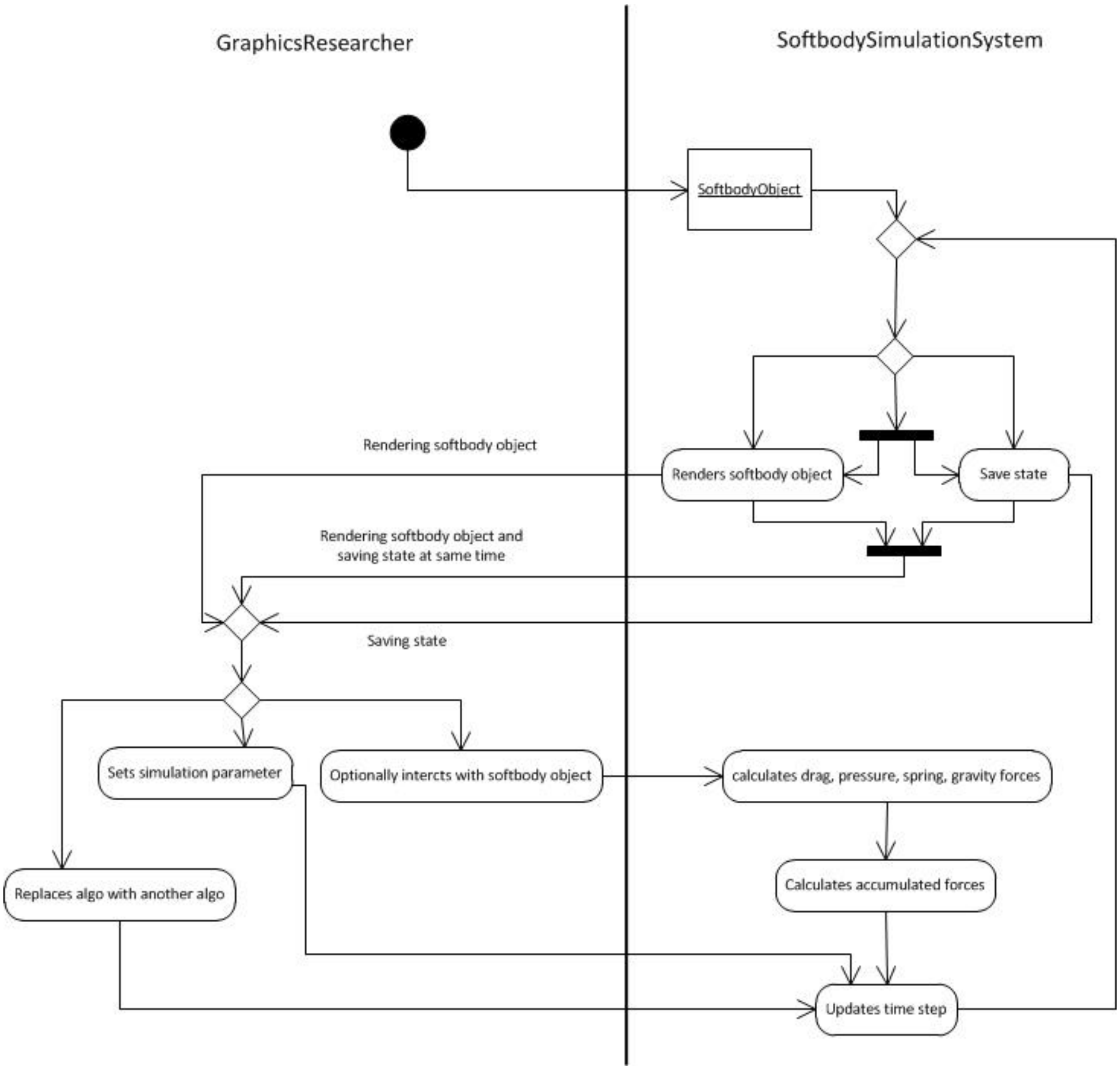

*Figure 6: activity diagram for SingleAlgo*

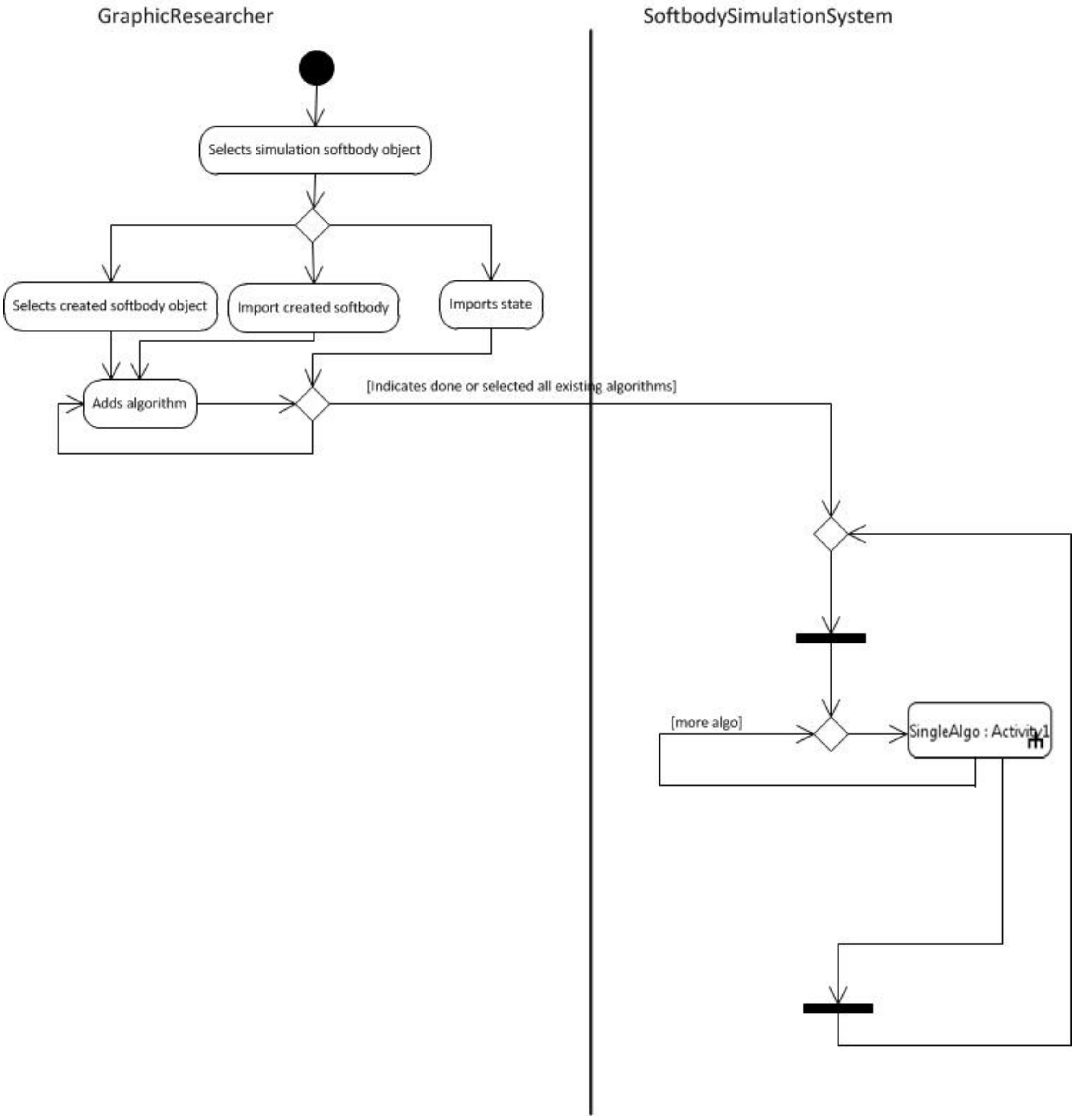

*Figure 7: activity diagram for Simulate Softbody Object*

As shown in figure7, when user launches the simulation execution for a softbody object, user can open a previously created softbody object, or import a created softbody object which is created by another system or import a state. User has option to add multiple algorithms for integration and collision detection. Then after the selection, and setting simulation parameters, the SingleAlgo is called in parallel for all chosen algorithms.

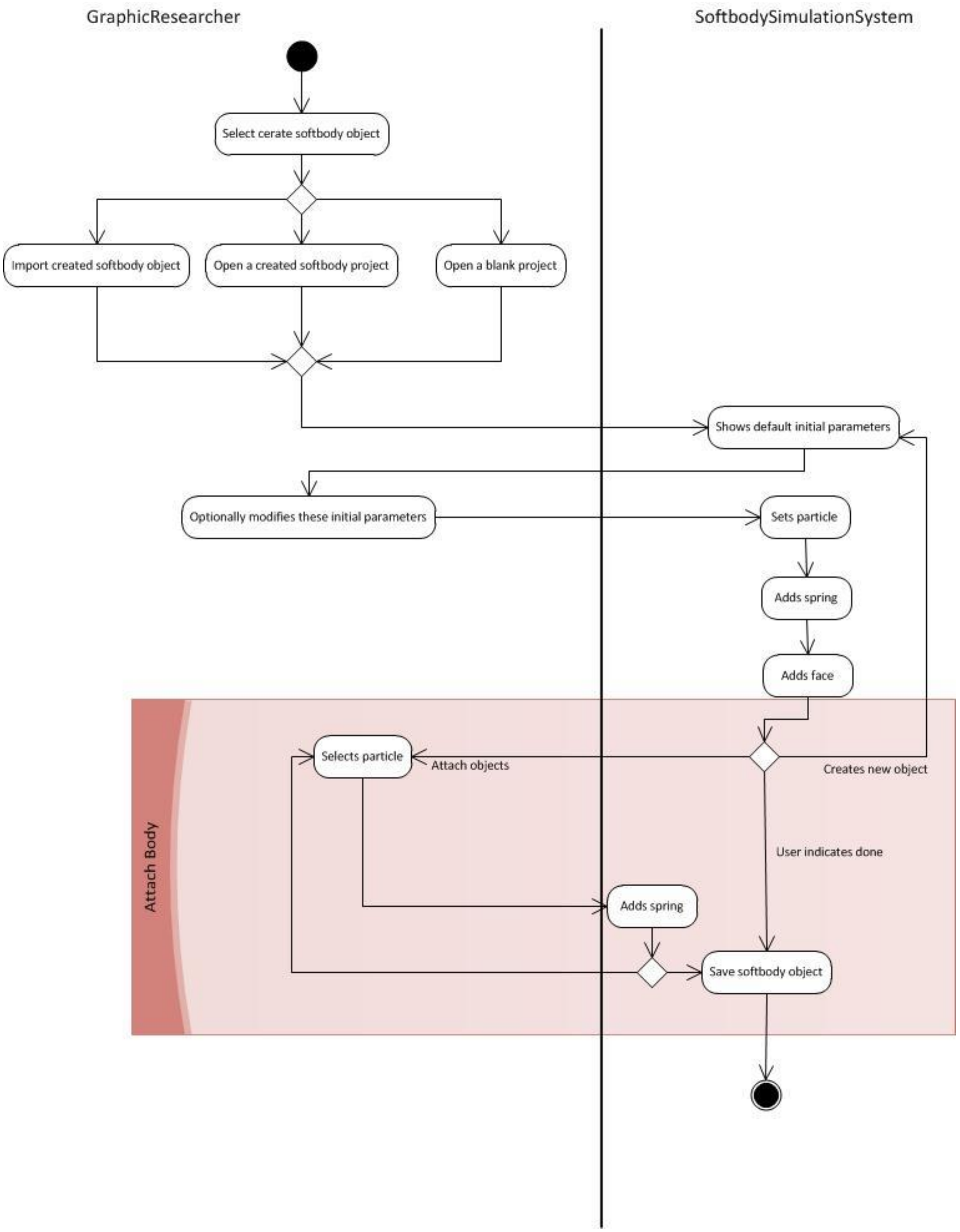

*Figure 8: activity diagram for Create Softbody Object*

# 4. Test cases

In this section, the test cases for use cases Simulate Softbody Object (UC01) and Create Softbody Object (UC02), one test case for the main scenario and two test cases for the alternative scenarios are shown.

## *4.1 Test cases for use case Simulate Softbody Object (UC01)*

| | |
|---|---|
| Test Case ID | TC1-UC01 |
| Title | Apply Algorithm |
| Requirement | F031 Apply integration algorithm |
| Type | Regular |
| Settings | N/A |
| Preconditions | User created softbody object before. |
| Description | User selects an integration algorithm. |
| Expected Results | Softbody object either responses to user's action or behaves as its idle behavior. |
| Test Case ID | TC2-UC01 |
| Title | Set Simulation Parameter |
| Requirement | F051 Set simulation parameter at run time |
| Type | Regular |
| Settings | N/A |
| Preconditions | Simulation is running. |
| Description | User changes the number of particle for a softbody object. |
| Expected Results | System renders softbody object with new properties and continues simulation. |
| Test Case ID | TC3-UC01 |
| Title | Change Algorithm |
| Requirement | F032 Change algorithm at run time |
| Type | Regular |
| Settings | N/A |
| Preconditions | Simulation is running. |
| Description | User changes the integration algorithm in the middle of simulation. |
| Expected Results | System renders softbody object according to the new algorithm within the new algorithm's interval time. |

### *4.2 Test cases for use case Create Softbody Object (UC02)*

| | |
|---|---|
| Test Case ID | TC1-UC02 |
| Title | Crete Softbody Object |
| Requirement | F012 Set initial parameter, F013 add spring and face |
| Type | Regular |
| Settings | N/A |
| Preconditions | There is no specific precondition. |
| Description | – User sets the properties of a softbody object like its dimension, number of particles and mass of each particle. |
| Expected Results | System creates a softbody object in a new project. |
| Test Case ID | TC2-UC02 |
| Title | Import Object |
| Requirement | F071 import a created object |
| Type | Regular |
| Settings | N/A |
| Preconditions | There is no specific precondition. |
| Description | – User selects the path that created object exists. |
| Expected Results | System loads the project. |
| Test Case ID | TC3-UC02 |
| Title | Attach Objects |
| Requirement | F014 Attach softbody object to another object |
| Type | Regular |
| Settings | N/A |
| Preconditions | There must be at least two objects to attach to each other. |
| Description | – User selects two sets of particle from two objects.<br>– System adds spring between particles of two set. So the head and tail of spring are not the particle of the same objects. |
| Expected Results | System creates a multi-layer object. |

# 5. Traceability

## *5.1 User needs versus features*

This table shows that each need is captured by which features of the system. For a full list of features refer to section 4.3 of vision document.

| | F011 | F012 | F013 | F014 | F021 | F022 | F023 | F024 | F025 | F026 | F027 | F031 | F032 | F033 | F041 | F051 | F061 | F071 | F072 | F073 | F081 | F082 | F091 | F101 | F111 | F112 |
|---|---|---|---|---|---|---|---|---|---|---|---|---|---|---|---|---|---|---|---|---|---|---|---|---|---|---|
| N01 | X | X | X | X | | | | | | | | | | | | | | | | | | | | | | |
| N02 | | | | | X | X | X | X | X | X | X | | | | | | | | | | | | | | | |
| N03 | | | | | | | | | | | | X | X | X | | | | | | | | | | | | |
| N04 | | | | | | | | | | | | | | | X | | | | | | | | | | | |
| N05 | | | | | | | | | | | | | | | | X | | | | | | | | | | |
| N06 | | | | | | | | | | | | | | | | | X | | | | | | | | | |
| N07 | | | | | | | | | | | | | | | | | | X | X | X | | | | | | |
| N08 | | | | | | | | | | | | | | | | | | | | | X | X | | | | |
| N09 | | | | | | | | | | | | | | | | | | | | | | | X | | | |
| N10 | | | | | | | | | | | | | | | | | | | | | | | | X | | |
| N11 | | | | | | | | | | | | | | | | | | | | | | | | | X | X |

*Table 5: traceability among the needs and features*

## *5.2 Features versus use cases*

Each use case is a set of steps to get an objective and it captures one to many features of the system. The following table shows that each use case contains which features of the system. As shown in this table, the feature F101 (adapting different classes and functions of system as API to another system) is not captured by any use cases.

| | F011 | F012 | F013 | F014 | F021 | F022 | F023 | F024 | F025 | F026 | F027 | F031 | F032 | F033 | F041 | F051 | F061 | F071 | F072 | F073 | F081 | F082 | F091 | F101 | F111 | F112 |
|---|---|---|---|---|---|---|---|---|---|---|---|---|---|---|---|---|---|---|---|---|---|---|---|---|---|---|
| UC01 | | | | | | | | | | | | X | X | | | X | | | X | | | | X | | | |
| UC02 | X | X | X | | | | | | | | | | | | | | | X | | | | | | | | |
| UC03 | | | | X | | | | | | | | | | | | | | | | | | | | | | |
| UC04 | | | | | | X | X | X | X | X | | | | | | | | | | | | | | | | |
| UC05 | | | | | X | | | | | | | | | | | | | | | | | | | | | |
| UC06 | | | | | | | | | | | | | | | X | | | | | | | | | | | |
| UC07 | | | | | | | | | | | | | | | | | | | | | | | | | | X |
| UC08 | | | | | | | | | | | | | | | | | | | | | | | | | X | |
| UC09 | | | | | | | | | | | | | | | | | | | | | X | X | | | | |
| UC10 | | | | | | | | | | | | | | | | | | X | | | | | | | | |
| UC11 | | | | | | | | | | | X | | | | | | | | | | | | | | | |
| UC12 | | | | | | | | | | | X | | | | | | | | | | | | | | | |
| UC13 | | | | | | | | | | | | | | X | | | | | | | | | | | | |
| UC14 | | | | | | | | | | | | | | | | | | | X | | | | | | | |
| UC15 | | | | | | | | | | | | | | | | | | | | X | | | | | | |
| UC16 | | | | | | | | | | | | | | | | | X | | | | | | | | | |

*Table 6: traceability among use cases and features*

## *5.3 Features versus supplementary requirements*

In the following table, the rows of the table are the features determined in section 4.3 of vision document and the columns are the non-functional requirements which are specified in supplementary document. This table shows that each non-functional requirement is related to which feature. For example, the features like providing a default softbody object (F011), creating environment (F021), rendering the softbody object (F022), setting the simulation parameters via a GUI (F051), and whatever makes the system easy to work affects the usability.

| | NF01 | NF021 | NF022 | NF03 | NF04 | NF05 | NF06 | NF07 |
|---|---|---|---|---|---|---|---|---|
| F011 | X | | | | | | | |
| F012 | X | X | | X | | | | X |
| F013 | | | | | | | | |
| F014 | | | | | | | | |
| F021 | X | | | | | | | |
| F022 | X | | | | | | | |
| F023 | | | | | | | | |
| F024 | | | | | | | | |
| F025 | | | | | | | | |
| F026 | | | | | | | | |
| F027 | | | | | | | | |
| F031 | | X | X | X | | | X | X |
| F032 | | X | X | X | | | X | X |
| F033 | | | | | | | | |
| F041 | X | | | | | | | |
| F051 | X | X | | X | | | | X |
| F061 | | | | | | | | |
| F071 | | | | | | | | |
| F072 | | | | | | | | |
| F073 | | | | | | | | |
| F081 | | | | | | X | | |
| F082 | | | | | | X | | |
| F091 | X | | | | | | | |
| F101 | | | | | | X | | |
| F111 | X | | | | | | | |
| F112 | X | | | | | | | |

*Table 7: traceability between features and non-functional requirements.*

### *5.4 Features versus supplementary requirements*

In the following table, the rows of the table are the fully dressed use cases defined in section 2 of use case document and the columns are the test cases that are defined in section 4 of this document. This table shows that each test case is for which use case.

| | TC1-UC01 | TC2-UC01 | TC3-UC01 | TC1-UC02 | TC2-UC02 | TC3-UC02 |
|---|---|---|---|---|---|---|
| UC01 | X | X | X | | | |
| UC02 | | | | X | X | X |

*Table 8: traceability between test cases and use cases.*

## References

[1] M. Song, "Dynamic deformation of uniform elastic two-layer objects," Master's thesis, Department of Computer Science and Software Engineering, Concordia University, Montreal, Canada, Aug. 2007, ISBN: 978-0-4943-4780-5.

[2] M. Song and P. Grogono, "Deriving software engineering requirements specification for computer graphics simulation systems through a case study", International conference on Information Sciences and Interaction Sciences (ICIS), 2010, pp. 285-291.

[3] M. Song and P. Grogono, "A framework for dynamic deformation of uniform elastic two-layer 2D and 3D objects in OpenGL," in Proceedings of C3S2E'08. ACM, May 2008, pp. 145–158, ISBN 978-1-60558-101- 9.

[4] M. Song and P. Grogono, "An LOD control interface for an OpenGL-based softbody simulation framework," in Innovations and Advances in Computer Sciences and Engineering, Proceedings of CISSE'08, T. Sobh, Ed. Springer Netherlands, Dec. 2008, pp. 539–543, published in 2010.

[5] Axel van Lamsweerde, "Requirements Engineering: From System Goals to UML Models to Software Specifications", chapter 3, Wiley, 2009. ISBN: 978-0-470- 1270-3.

[6] http://anybody.software.informer.com/ [online; accessed 01 April 2013].

[7] http://softimage.ru/wiki/index.php/[online; accessed 01 April 2013].

[8] https://en.wikipedia.org/wiki/Haptic_technology [online; accessed 25 July 2013].

[9] J. F. O'brien, C. Shen, and C. M. Gatchalian, "Synthesizing Sounds from Rigid-Body Simulations", Proceedings of the 2002 ACM SIGGRAPH/Eurographics symposium on Computer animation, 2002, pp. 175-181.

[10] IEEE- SA Boards, "IEEE Recommended Practice For Software Requirements Specification", 2009.

## Interview

**Major parts of this document obtained from Ms. Song Master Thesis [1] and refined after scheduling meeting with Mr. Surguei A Mokhov and Ms. Miao Song.**

**Part I: Establishing the customer or user Profile:**
Name: Ms. Song
Company: Concordia University
Job title: Graphics Researcher

---

**What are your key responsibilities?** Graphics researcher

**What outputs do you produce?** A softbody simulation system which is able to interact with objects outside the system at real time.

**For whom?** The core system was done under the supervision of Dr. Grogono as Master thesis in Concordia University.

**How is success measured?** By accuracy of created softbody object, performance and stability of algorithm that is used for simulating softbody object at real time.

**Which problems interfere with your success?**

- Bounded by available algorithms.
- Complexity of other systems like Maya for creation of softbody object.
- Interacting with softbody object at real time.
- Adding algorithms at the run time for comparison.
- Import created bodies from other system.
- Attaches the created bodies, like rigidbody to a softbody.

**What, if any, trends make your job easier or more difficult**? There are different tools which the best one is Maya .They are not complete as for the simulation, it can considers pre- defined parameters and user cannot interact with them or change them at simulation time.

**Part II: assessing the problem**

For which/ application type/problems do you lack good solution? Interacting with user at real time.

For which problem, ask the following questions.

- Why does this problem exist? Because systems like Maya does simulation with predefined parameters.
- How do you solve it now? No solution with Maya.
- How would you like to solve it? By interacting with user by devices like mouse, haptic devices.

**Part III: understanding the user environment**

**Who are users?** There are different types of users that use the system. In general, the users for Softbody simulation system range from computer graphics researcher and game developers to kids, surgeon who interacts with the Softbody. But the jelly fish as application was used as case study. Other research students can also use the core system and extend it for their purposes.

**What is their educational background?** The users who create the softbody know the graphics computer. But for end users who interact with the simulated softbody, it differs.

**What is their computer background?** Above mentioned answer.

**Are users experienced with this type of application?** May not

**Which platforms are in use?** Windows as operating system at first but in later versions it works Linux, Unix and OS X.

**What are your expectations for usability of the product?** It should be easy for graphics researcher to enter the parameters for the creation of the Softbody like the number of particles for the Softbody as LOD and the created softbody should be easy for end users to interact with.

**What are your expectations for training time?** A system is easy for end user to use it. A short training would be enough.

**What kinds of user help (for example, hard copy and online documentation) do you need?** Both can be helpful.

**Part IV: recap for understanding**
You have told me:
The system should create the softbody object easily and allows for selecting the algorithms at the time of simulation while it allows interacting with softbody. Moreover the system should be able to add new algorithms to the system and attach the previously created bodies to each other.
Does this adequately represent the problems you are having with your exciting solution?
What, if any, other problems are you experiencing? No

**Part VII: Assessing your opportunity**

**Who in your organization needs this application?** Graphics computer researchers to create the softbody object or other students as researchers use this system as the core system and extend it.

**Part VIII: Assessing the Reliability, Performance, and Support Needs**

**What are your expectations for reliability**? There must be a close approximation between the created softbody object and real object both in shape and behavior and response to the actions in real time. And the system shouldn't fail frequently and the repair time should be reasonable.

**What are your expectations for performance?** The system should be fast enough to do the computations. Although being fast enough in computations and having real time response by softbody object is in contradict with accuracy which is related to the LOD in softbody object. So depending on the type of application, there is a trade of between accuracy and performance.

**Will you support the product, or will others support it?** This application is for educational purposes.

**Do you have special needs for support?** No

**What about maintenance and service access?** The system is supported by myself.

**What are security requirements?** There is no specific security requirement for the core softbody simulation system but regarding to the application, it can be considered as well.

**How will the software be distributed?** Currently, it is open source software.

**Part V: The analyst's input on the customer's problem**

What if the graphics researcher doesn't know about the physics rules and forces?
Is this a real problem? No
How do you currently solve the problem? The current rules are simple ones and the applied forces can be gravity force, spring, or drag force. Moreover, these physical rules are considered automatically when the simulation is started and user does not need to know which physical rules need to apply.
Are those physics rules enough? May not. Depending on the object and the application.
How would you rank solving these problems in comparison to others you've mentioned? Low

**Part VI: Assessing your solution (if applicable)**
What if any could:
How would you rank the importance of these? N/A

**Part IX: other requirements**
Are there any legal, regulatory, or environmental requirements or other standards that must be supported? The system should be able to interact with input devices like haptic devices.
Can you think of any other requirements we should know about? For doing computation, GPU as the processing unit is needed.
**Part X: wrap-up**
Are there any questions I should be asking you? No If I need to ask follow-up questions, may I email you? Would you like to participate in a requirements review? Yes

**Part XI: The analyst's summary**

After the interview, and while the data is still fresh in your mind, summarized the three highest priority needs or problems identified by this user/customer.

1- Adding algorithm at run time.
2- Interact with created softbody object at real time.
3- Enter the parameters for creation of softbody easily.